\documentclass[preprint,showpacs,preprintnumbers,amsmath,amssymb,nofootinbib]{revtex4}

\usepackage{etex}

\usepackage{amsthm,amscd,amsbsy,array}
\usepackage{bm}
\usepackage{soul} 

\usepackage{graphics,graphicx,xcolor}

\usepackage{soul} 

\usepackage[colorlinks=true, pdfstartview=FitV, linkcolor=blue, citecolor=blue, urlcolor=blue]{hyperref} 

\newcommand{\red}{\textcolor{red}}  
\newcommand{\blue}{\textcolor{blue}}

\newcommand{\gb}{\colorbox{green}}

\newenvironment{redtext}{\color{red}}{\ignorespacesafterend} 
\newenvironment{bluetext}{\color{blue}}{\ignorespacesafterend}

\newcommand{\bblue}{\begin{bluetext}} 
\newcommand{\eblue}{\end{bluetext}} 
\newcommand{\bred}{\begin{redtext}}
\newcommand{\ered}{\end{redtext}}

\numberwithin{equation}{section}

\let\ssection=\section
\renewcommand{\section}{\setcounter{equation}{0}\ssection}

\newcommand{\cA}{{\mathcal{A}}}

\newcommand{\bb}{{\bf b}}

\newcommand{\bone}{\boldsymbol{1}}

\newcommand{\bc}{{\mathbf{c}}}

\newcommand{\cC}{{\mathcal{C}}}

\renewcommand{\d}{\mathrm{d}}

\newcommand{\diag}{\mathrm{diag}}

\newcommand{\rE}{{\mathrm{E}}}

\newcommand{\cB}{{\mathcal{B}}}

\newcommand{\rg}{\mathrm{g}}

\newcommand{\Id}{\mathrm{Id}}

\newcommand{\bk}{\mathbf{k}}

\newcommand{\rO}{{\mathrm{O}}}

\newcommand{\bp}{{\bf p}}

\newcommand{\br}{{\bm{r}}}
\newcommand{\bx}{{\bm{x}}}

\newcommand{\bbR}{\mathbb{R}}

\newcommand{\Tr}{\mathrm{Tr}}

\newcommand{\bX}{{\bf X}}

\def\smallover#1/#2{\hbox{$\textstyle\frac{#1}{#2}$}} %

\def\Ort{{\rm O}}

\def\bp{{\bm{p}}}

\def\parag{\hfil\break} 
\def\kikezd{\parag\underbar}

\def\benu{\begin{enumerate}}
\def\eenu{\end{enumerate}}
\def\beq{\begin{equation}}
\def\eeq{\end{equation}}
\def\beqa{\begin{eqnarray}}
\def\eeqa{\end{eqnarray}}

\def\barray{\left(\begin{array}}
\def\earray{\end{array}\right)}
\def\barraynb{\begin{array}}
\def\earraynb{\end{array}}

\def\Ort{{\rm O}}
\def\IR{{\mathbb{R}}} 

\def\?{\quad{\gb{\fbox{\texttt{?}}\;}}\quad}
\def\p{{\partial}}

\def\v0{\mathbf{0}}

\def\Rarrow{{\quad$\Rightarrow$\quad}}

\def\beq{\begin{equation}}
\def\eeq{\end{equation}}
\def\bea{\begin{eqnarray}}
\def\eea{\end{eqnarray}}

\def\p{\partial}

\def \p{{\partial}}

\def\6{\partial}
\def\7{\tilde}
\def\8{\widehat}
 \def\bx{{\bf x}}

\def\G11{\Gamma_{11} }

\newcommand{\const}{\mathop{\rm const.}\nolimits}
\newcommand{\half }{\frac{1}{2}}

\def\smallover#1/#2{\hbox{$\textstyle\frac{#1}{#2}$}} %
\def\smallcirc{{\raise 0.5pt \hbox{$\scriptstyle\circ$}}}
\def\2{{\smallover1/2}}

\let\ssection=\section
\renewcommand{\section}{\setcounter{equation}{0}\ssection}

\begin{document} 

\preprint{\texttt{arXiv:1702.08284v3 [gr-qc]}}

\title{Carroll symmetry of plane gravitational waves
\\[6pt]
}

\author{
C. Duval$^{1}$\footnote{
mailto:duval@cpt.univ-mrs.fr},
G. W. Gibbons$^{2,3,4}$\footnote{
mailto:G.W.Gibbons@damtp.cam.ac.uk},
P. A. Horvathy$^{3,5}$\footnote{mailto:horvathy@lmpt.univ-tours.fr},
P.-M. Zhang$^{5}$\footnote{e-mail:zhpm@impcas.ac.cn}
}

\affiliation{
$^1$Aix Marseille Univ, Universit\'e de Toulon, CNRS, CPT, Marseille, France
\\
$^2$D.A.M.T.P., Cambridge University, U.K.
\\
$^3$Laboratoire de Math\'ematiques et de Physique
Th\'eorique,
Universit\'e de Tours,
France
\\
$^4$LE STUDIUM, Loire Valley Institute for Advanced Studies, Tours and Orleans France
\\
$^5$Institute of Modern Physics, Chinese Academy of Sciences, Lanzhou, (China)
}

\date{\today}


\begin{abstract} 
The well-known 5-parameter isometry group of
plane gravitational waves in $4$ dimensions is identified  as L\'evy-Leblond's Carroll group  in $2+1$ dimensions with no rotations.
 Our clue is that plane waves are Bargmann spaces into which Carroll manifolds can be embedded. We also comment on the scattering of light by a gravitational wave and calculate its electric permittivity considered as an impedance-matched metamaterial.
\end{abstract}

\maketitle


\section{Introduction}\label{Intro}

A gravitational plane wave is the $4$-manifold, $\bbR^4$ globally, endowed with the Lorentz metric \cite{BaJeRo,BoPiRo,EhlersKundt,Sou73,exactsol,Gibb75,Garriga,Torre}, 
\beq
\rg =\delta_{ij}\,dX^idX^j+2dUdV+K_{ij}(U){X^i}{X^j}\,dU^2,
\label{planewave}
\eeq
where the symmetric and traceless matrix
$K(U)=\left(K_{ij}(U)\right)$ characterizes the profile of the wave. Here $U$ and $V$ are light-cone coordinates, whereas $\bX = (X^1,X^2)$ parametrizes the transverse plane which carries the flat Euclidean metric $d\bX^2=\delta_{ij}\,dX^idX^j$. This metric is a Brinkmann pp-wave metric \cite{Bri}. Since 
the only non-vanishing curvature components are 
 $R^i_{\,UjU}=-R^V_{\,ijU}
 =-K_{ij}$, it is Ricci-flat. 
In~$3+1$ dimensions, (\ref{planewave}) is the general form of a Ricci-flat Brinkmann metric. 
 
In this paper we identify, following \cite{Sou73}, the isometry group  of  the gravitational wave (\ref{planewave}) we denote here by $C$,
 as the \emph{Carroll group in $2+1$ dimensions with broken rotations}. 

Let us recall that the Carroll group, $C(d+1)$, has originally been introduced as an unusual contraction, $c\to0$, of the Poincar\'e group $\rE(d,1)$ \cite{Leblond, SenGupta,Carrollvs}.
 It acts on $(d+1)$-dimensional flat ``Carroll space-time", $\cC^{d+1}$, with coordinates $(\bx,v)$, according to
\beq
\bx\to A\,\bx +\bc,
\qquad
v \to v -\bb\cdot{}A\,\bx + f,
\label{Carrolla}
\eeq
where $A\in\Ort(d)$ and $\bc\in\IR^d$ represent spatial orthogonal transformations and translations; $\bb\in\IR^d$ and $f\in \IR$ generate ``Carrollian boosts''  and translations of ``Carrollian time'', $v$, respectively.
Let us recall that the flat Carroll space-time, $\cC^{d+1}\cong\bbR^{d+1}$, is endowed with the degenerate metric $\gamma=d\bx^2$ with kernel spanned by the constant vector field $\xi=\partial_v$, and is equipped with its flat affine connection. The Carroll group, $C(d+1)$, is thus the group of all affine transformations of ``space-time'' which leave $\gamma$ and $\xi$ invariant.


Long considered as a sort of mathematical curiosity \cite{Ancille, Carrollvs, BeGoLo}, the Carroll group has reemerged more recently in brane-dynamics 
 \cite{GiHaYi,thoughts}, for the BMS group \cite{DGH-BMS}, in string theory \cite{CaGoPo,DGH-BMS,Bag}, and in non-relativistic gravitation \cite{BeGoRoRoVe}.

We find it useful to record our notations:
$(\bX,U,V)$ denote ``Brinkmann'' coordinates which appear in~(\ref{planewave}), whereas  $(\bx,u,v)$ in (\ref{BJRcoord}) are Baldwin-Jeffery-Rosen (BJR) coordinates \cite{BaJeRo,Sou73}, convenient for determining both the isometries and the geodesics \cite{Sou73}. In Minkowski space-time, $K=0$,  both coordinates reduce to  the usual light-cone coordinates
$(\br,t,s)$.

\section{The Carroll group and Bargmann symmetries}\label{planewaves}

Our clue is that the space-time with metric (\ref{planewave}) is also a \emph{Bargmann manifold} \cite{DBKP,DGH91} in that it carries a null, covariantly constant, vector field, namely $\xi=\p_V$. 

Bargmann spaces are convenient tools to study non-relativistic dynamics in one lower dimension \cite{DBKP,DGH91,Eisenhart,Cariglia}~:
the quotient of a $(d+1,1)$-dimensional Bargmann space by the foliation generated by $\xi$ carries a $(d+1)$-dimensional Newton-Cartan structure  \cite{DBKP}; $U$ in (\ref{planewave}), can be viewed as \emph{non-relativistic time} and $-\frac{1}{2}K_{ij}(U){X^i}{X^j}\,$ as a time dependent quadratic scalar potential.  
Non-relativistic motions in Newton-Cartan spacetime are projections of null geodesics in  extended Bargmann space-time \cite{Eisenhart,DBKP,DGH91}; the $\xi$-preserving isometries of the latter project to non-relativistic Galilean symmetries.

 The metric (\ref{planewave}) describes, in Bargmann language, the Eisenhart lift of a (possibly anisotropic) planar oscillator with time dependent frequencies \cite{DBKP,DGH91,Eisenhart,Cariglia,BDP}; its $\xi$-preserving isometries span the [centrally extended] Newton-Hooke group [without rotations in the anisotropic case] \cite{ZHAGK};  when $K=0$, the latter goes over to the [centrally extended]  Galilei group (also referred to as the Bargmann group in $(d+1,1)$ dimension) \cite{DBKP,DGH91}. The latter is conveniently represented by the matrices 
\beq
\left(
\begin{array}{cccc}
A&\bb&0&\bc\\
0&1&0&e\\
-\bb^{\dagger}{}A&-\half\bb{}^2&1&f\\
0&0&0&1
\end{array}
\right),
\label{Barggroup}
\eeq
where $A\in\rO(d)$; $\bb,\bc\in\IR^d$; and $e,f\in\IR$ stand for a Galilei time translation and for a ``Carroll time'' translation, respectively. The superscript ${}^{\dagger}$ means  transposition. The Bargmann group (\ref{Barggroup}) acts affinely by matrix multiplication on
Bargmann extended spacetime $\bbR^{d+1,1}$ parametrized by $(\br,t,s)$, according to
\begin{equation}
\br\to{}A\br+\bb t+\bc,
\qquad
t\to{}t+e,
\qquad
s\to s-\bb\cdot{}A\br-\half\bb^2{}t+f.
\label{BargAction}
\end{equation}

Recall that  flat Bargmann space is endowed with the metric $\rg=\d\br^2+2dt\,ds$, and the null, constant, vector field $\xi=\partial_s$. The null hyperplane $\cC_{t_0}^{d+1}$ defined by  $t=t_0=\const$ is a  \emph{Carroll spacetime}  \cite{Carrollvs,DGH-BMS} as it carries a twice-symmetric, semi-positive ``metric'', $d\br^2=\rg_{\vert{}t=t_0}$, whose kernel is generated by $\xi$. The restriction of the Levi-Civita connection of the Bargmann metric is then a 
distinguished Carroll connection on $\cC_{t_0}^{d+1}$ \cite{DGH-BMS}, coordinatized by $\bx=\br$ and $v=s$. 
The action (\ref{Carrolla}) of the Carroll group is readily recovered by putting $e=0$ in (\ref{BargAction}) and performing an easy redefinition of space and ``Carroll time'' translations. The Carroll group, $C(d+1)$ is hence the subgroup of the Bargmann group (\ref{Barggroup}), defined by
$
e=0.
$
It is therefore a subgroup of all isometries, i.e., of the Poincar\'e group, $\rE(d+1,1)$, in the flat case. From now on, we set $d=2$ again.

\section{Isometries of plane gravitational waves}\label{isos}

 
Our first step in identifying the isometries of the plane wave (\ref{planewave}) is 
to recast the metric in a new coordinate system $(\bx,u,v)$ \cite{BaJeRo,Sou73}, which allows us to determine the isometry group in terms of elementary functions, see eqn. (\ref{genCarr}) below.  
This can be achieved \cite{Gibb75,exactsol,Garriga} with the help of a change of coordinates $(\bX,U,V)\to(\bx,u,v)$, namely
\beq
{\bX} =P(u)\bx,
\qquad
U=u,
\qquad 
V=v-\frac{1}{4}\bx\cdot\dot{a}(u)\bx\,,
\label{BJRcoord}
\eeq
where $a(u)=P(u)^\dagger{}P(u)$ for some non-singular $2\times2$ matrix $P(u)$\footnote{
The square root, $P$, of the matrix $a>0$ is not uniquely defined since $RP$ with $R(u)\in\rO(2)$ is another one. However, eqns (\ref{SL+cond})  guarantee that $\dot{R}=0$, i.e., that $P$ is merely defined up to $\rO(2)$. Conditions (\ref{SL+cond}) implies that
$ 
K=\ddot{P}P^{-1}
$ 
in (\ref{planewave}) is indeed symmetric.}. 
The change of coordinates allows us to present the metric (\ref{planewave}) in the form
\beq
\rg=a_{ij}(u)\,dx^idx^j+2du\,dv,
\label{nopotform}
\eeq
\textit{provided}  the matrix $P(u)$ is a solution of the joint system
\beq
\ddot{P}= KP,
\qquad
P^\dagger\dot{P}=\dot{P^\dagger}P.
\label{SL+cond}
\eeq
Since  
\beq
K=\half{}P\left(\dot{L}+\half{L}^2\right)P^{-1},
\qquad
L=a^{-1}\dot{a},
\label{KfromP}
\eeq
the Ricci flatness of the metric (\ref{nopotform}), namely $\Tr(K)=0$, is given by the equation \cite{BaJeRo,Sou73},
\beq
\Tr\left(\dot{L}+\half{}L^2\right)=0.
\label{Ricci=0}
\eeq\vskip-5mm
\kikezd{Isometries}.
 We now claim that the (generic) isometries of (\ref{nopotform}) form  a $5$-dimensional Lie group. Following Souriau \cite{Sou73},  in terms of the BJR  $(\bx,u,v)$, the latter acts on space-time as, 
\beq
\bx\to\bx+H(u)\bb+\bc,
\qquad
u\to u,
\qquad
v\to v-\bb\cdot\bx - \2\bb\cdot{}H(u)\bb+f,
\label{genCarr}
\eeq
where $H$ is a (symmetric) $2\times 2$ matrix verifying $\dot{H}=a^{-1}$, that is,
\beq
H(u)=\int^u_{u_0}\!\!a(t)^{-1} dt.
\label{Hmatrix}
\eeq

Manifestly, $\bc\in\IR^2$ and $f$ generate transverse-space resp. null translations along the~$v$ coordinate \cite{Gibb75}. Moreover, the group law 
deduced from (\ref{genCarr}),
\beq
(\bb,\bc,f).(\bb',\bc',f')=
(\bb+\bb',\bc+\bc',f+f'-\bb\cdot\bc'),
\label{grouplaw}
\eeq
is precisely that of the Carroll group, (\ref{Carrolla}) with no rotations, i.e. $A=\Id$. Thus the isometry group of the above gravitational wave  is indeed the group $C\subset{}C(2+1)$ of matrices
\beq
\left(
\begin{array}{cccc}
\bone&\bb&0&\bc\\
0&1&0&0\\
-\bb^{\dagger}&-\half\bb{}^2&1&f\\
0&0&0&1
\end{array}
\right).
\label{CarinB}
\eeq
The vector $\bb\in\IR^2$ generates, in particular, a Carroll boost acting as in (\ref{genCarr}), while $\bc\in\bbR^2$ and $f$ are Carrollian ``space-time'' translations, as before. Let us note that $C$ is  a normal subgroup of the Carroll group of $C(2+1)$, $C(2+1)/C\cong\rO(2)$.

It is worth stressing that while the way the Carroll group is implemented on an $u=\const$ null hypersurface \emph{does} depend on the (fixed) value of $u$ \cite{Torre},  
all these actions can be derived from the simple Carroll action at $u=u_0$ according to (\ref{genCarr})-(\ref{Hmatrix}).
The coordinate transformation (\ref{BJRcoord}) followed backwards yields the Carroll action in terms of the Brinkmann coordinates and we recover eqn. \# (3) in \cite{Torre}.

\kikezd{Geodesic motion}.
 As noted by Souriau  \cite{Sou73}, the conserved quantities associated with geodesic motions 
 are then readily determined  by Noether's theorem. Choosing, with no restriction,~$u$ as parameter, they are
\beq
\bp = a\,\dot{\bx},
\qquad
\bk=\bx-H\bp,
\qquad
m=1,
\label{CarCons}
\eeq
interpreted as \emph{linear momentum, boost-momentum and ``mass''} (unity in our para\-metri\-zation).  
In  BJR coordinates, the geodesics have then the remarkable explicit expression
\beq
\bx(u)=H(u)\,\bp+\bk,
\qquad
v(u)=-\half \bp\cdot H(u)\,\bp + e\,u+ d,
\label{CarGeo}
\eeq
where $e=\half\rg_{\mu\nu}\,\dot{x}^\mu\dot{x}^\nu$ is.  
(This constant is negative/zero/positive for timelike/null/spacelike geodesics). $d$ is an integration constant. 
The isometry group acts on the above constants of the motion as
\beq
(\bp,\bk,e,d)\to (\bp+\bb,\bk+\bc,e,d+f-\bb\cdot\bk)
\label{actiononCC}
\eeq
allowing us, in particular, to carry any geodesic to  a geodesic defined by  $\bp=\bk=0$ and $d=0$, which then becomes ``vertical'' \cite{Sou73},
$ \bx=0,
\; v=c\,u.
$ 
Conversely, any geodesic can be obtained by ``exporting a vertical one'' by an isometry~: the only group-invariant property of the trajectory is the sign of $e$.

\section{Classical examples}\label{Section:Examples}

Let us illustrate our procedure by simple examples.

\begin{enumerate}
\item 
The restriction of flat Minkowski space with metric $\rg=d\br^2+2dt\,ds$ to the constant ``time'' slice $t_0=0$ is  a Carroll manifold with $\xi=\partial_s$, upon which the
restriction $e=0$ of the Bargmann group (\ref{Barggroup}) acts, consistently with the Carroll action (\ref{Carrolla}), via (\ref{genCarr}) with
$ 
H(t)= (t-t_0)\,\Id\,.
$ 
The conserved quantities (\ref{CarCons}) take the familiar Galilean form. 
Shifting the basepoint, $t_0\to t_0'$,  merely shifts the transformation law of $v$ by a constant, namely
$
v \to v -\bb.\bx -\2\bb^2 (t_0'-t_0)+f.
$   

\item
For a less trivial example, consider, e.g.,
\begin{equation}
P(u)=
\chi(u)\,\diag\big(e^{\alpha(u)},e^{-\alpha(u)}\big)
\label{Pter}
\end{equation}
with $\alpha(u)$ some arbitrarily chosen function. The associated metric 
\begin{equation}
\rg=\chi^2\left[e^{2\alpha}(dx^1)^2+e^{-2\alpha}(dx^2)^2\right]+2du\,dv
\label{gbis}
\end{equation}
is Ricci-flat if 
\beq 
\ddot{\chi}+\dot{\alpha}^2\,\chi=0,
\eeq
see (\ref{Ricci=0}). 
The pp-wave profile is, in this case,
\begin{equation}
K_{ij}(U)X^iX^j=\half\cA(U)\left[(X^1)^2-(X^2)^2\right],
\qquad
\cA=\frac{2}{\chi^2}\,\frac{\ d}{du}\left(\chi^2\,\frac{d{\alpha}}{du}\right).
\label{HXX}
\end{equation}

From the mechanical point of view, this metric describes two uncoupled time-dependent harmonic oscillators, one attractive, the other repulsive, with opposite frequency-squares.

The metric (\ref{gbis}) is manifestly invariant against transverse space and advanced time-translations, $\bx\to\bx+\bc$ and  $v\to{}v + f$, respectively, 
while retarded time, $u$, is kept fixed. The orthogonal group, $\rO(2)$, is clearly broken since the spatial metric is not diagonal unless $\alpha(u)=0$. Carroll boosts act as in (\ref{genCarr}) with
$ 
H(u)=\int^u_{u_0}\!\!P(w)^{-2} dw,
$ 
where $P$ is as in (\ref{Pter}).

To have a toy example, let us
take, e.g., $\alpha(u)=u$. Then a Ricci-flat metric which is regular in the neighborhood of $u=0$ is given, for example, by $\chi(u)=-\cos u$, whose  profile is
\beq
\half
K_{ij}(U)\,X^iX^j=\tan U \left[(X^2)^2-(X^1)^2\right].
\label{uprofile}
\eeq
It describes a saddle-like surface with ``time''-dependent scale, depicted on Fig.\ref{saddles}, which shows the change of the profile when $u$ passes from negative to positive\footnote{For $u=\pm\pi/2$ $\chi=\det(a(u))^{1/4}$ vanishes. This is a general property which indicates a mere \emph{coordinate singularity} \cite{Sou73}.}.
\begin{figure}[h]
\begin{center}
\includegraphics[scale=.3]{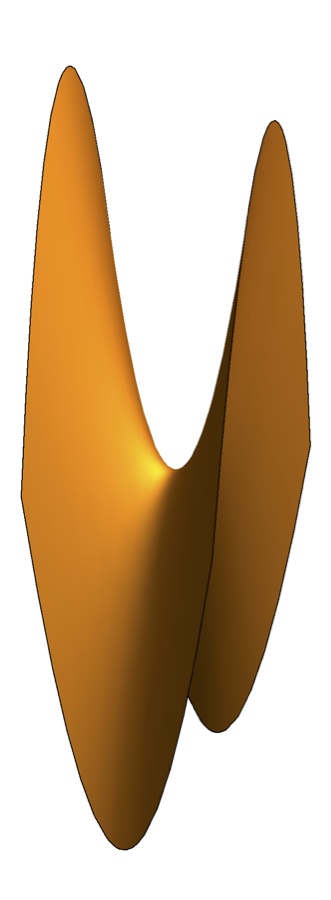}\;\,
\includegraphics[scale=.3]{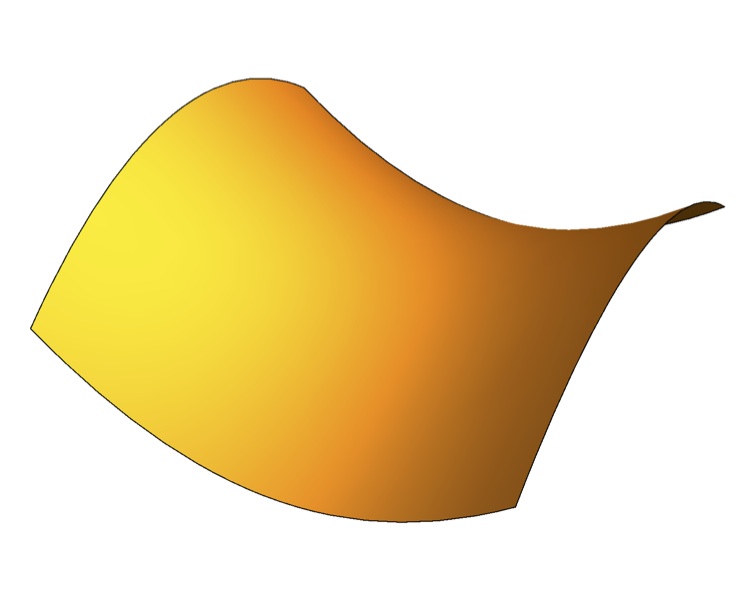}\;\,
\includegraphics[scale=.3]{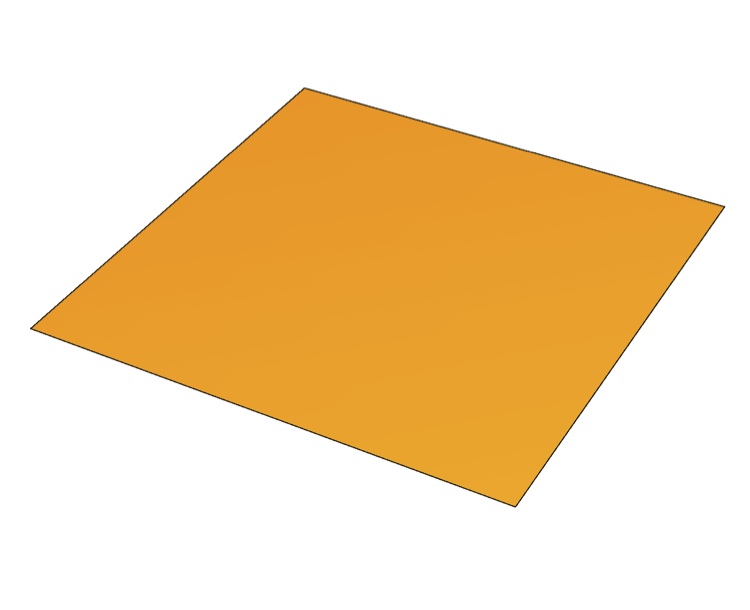}\;\,
\includegraphics[scale=.3]{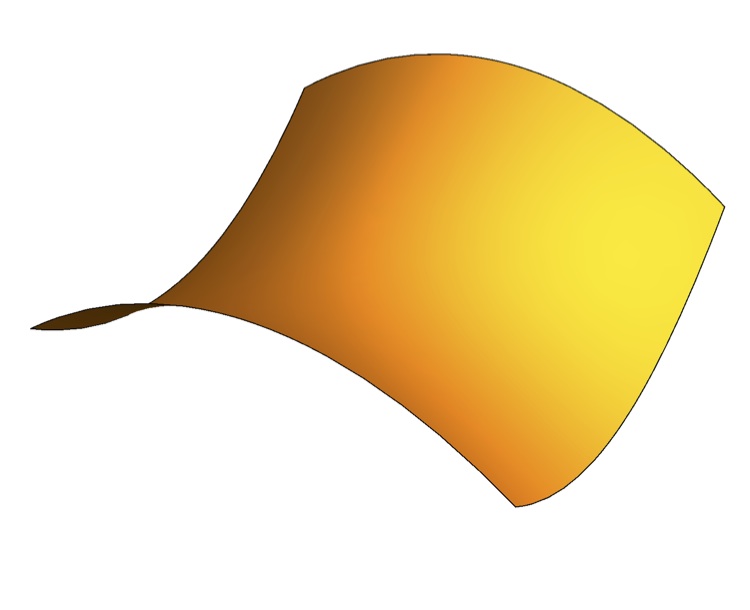}\;\,
\includegraphics[scale=.3]{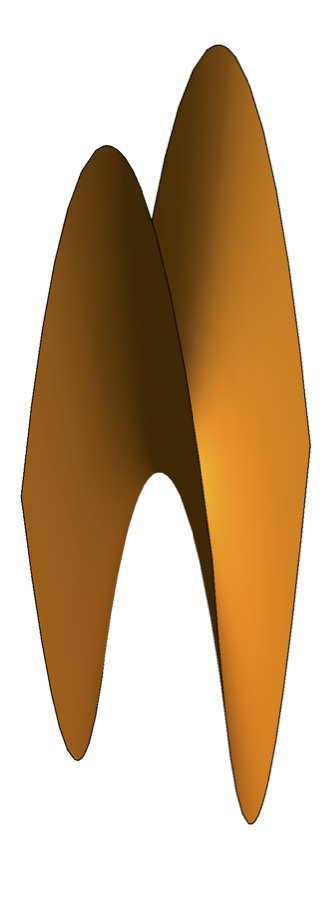}
\end{center}\vskip-8mm
\caption{{\it Wave profile (\ref{uprofile}) for $\alpha(u)=u$ for $u= -\pi/2 +0.1,\
u = -\pi/4,\
u = 0,\
u = \pi/4,\ 
u = \pi/2 -0.1$. For $u=0$ we get flat Minkowski space, as expected.}}
\label{saddles}
\end{figure}
\vskip2mm

The components of the matrix-valued function $H=\diag\big(H_{11}, H_{22}\big)$,
\beq
H_{11}(u)=\displaystyle\int_0^u\!\cos^{-2}w\,e^{- 2w}dw,
\quad
H_{22}(u)=\displaystyle\int_0^u\!\cos^{-2}w\,e^{+ 2w}dw
\label{Hfunc}
\eeq
 which rules both how Carroll boosts act, (\ref{genCarr}), and the evolution of geodesics, (\ref{CarGeo}), are plotted on Fig.\ref{Hfig}. For $u>0$ the component~$\red{H_{22}}$ increases rapidly while $\blue{H_{11}}$ is  damped. For $u<0$ the  behavior is the opposite, consistently with the change of profile shown on Fig.\ref{saddles}\,.\vskip-3mm
\begin{figure}[h]
\begin{center}
\includegraphics[scale=.35]{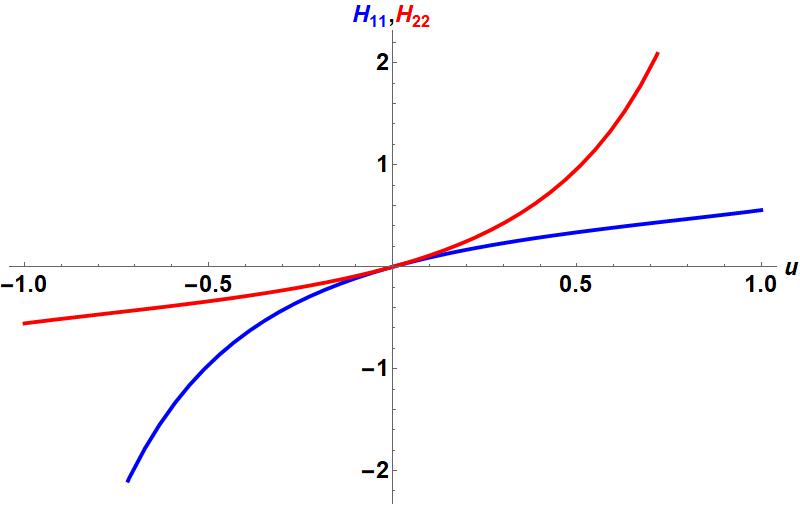}
\end{center}\vskip-8mm
\caption{ {\it The entries of the matrix-valued function $H(u)=\diag\big(H_{11},H_{22}\big)$ for $\alpha(u)=u$.}
} 
\label{Hfig}
\end{figure}

The evolution of a geodesic, shown on Fig.\ref{trajectory}a,
 is consistent with the profile change~: the repulsive and attractive directions are interchanged when $u$ changes sign. As said in sec. \ref{isos}, the curling trajectory could actually be ``straightened out'', see Fig.\ref{trajectory}b, by transforming $\bp\to0, \bk\to0, d\to0$ by a suitable action of the Carroll group, see (\ref{actiononCC}). 
\goodbreak
\begin{figure}[h]\vskip-3mm
\begin{center}
\includegraphics[scale=.22]{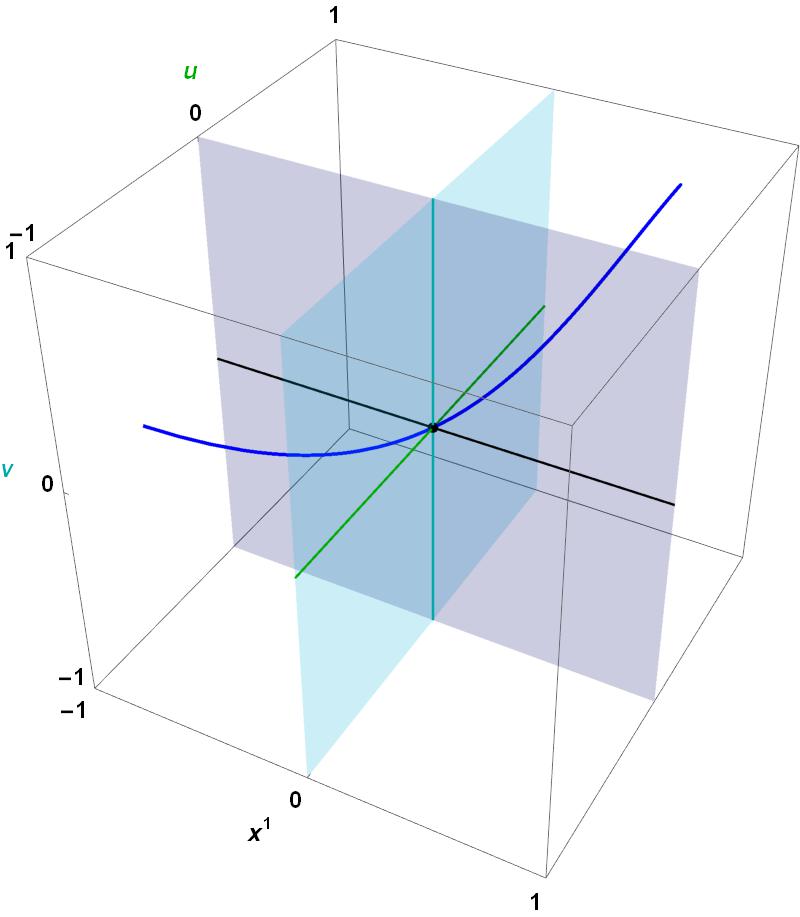}
\Rarrow
\includegraphics[scale=.22]{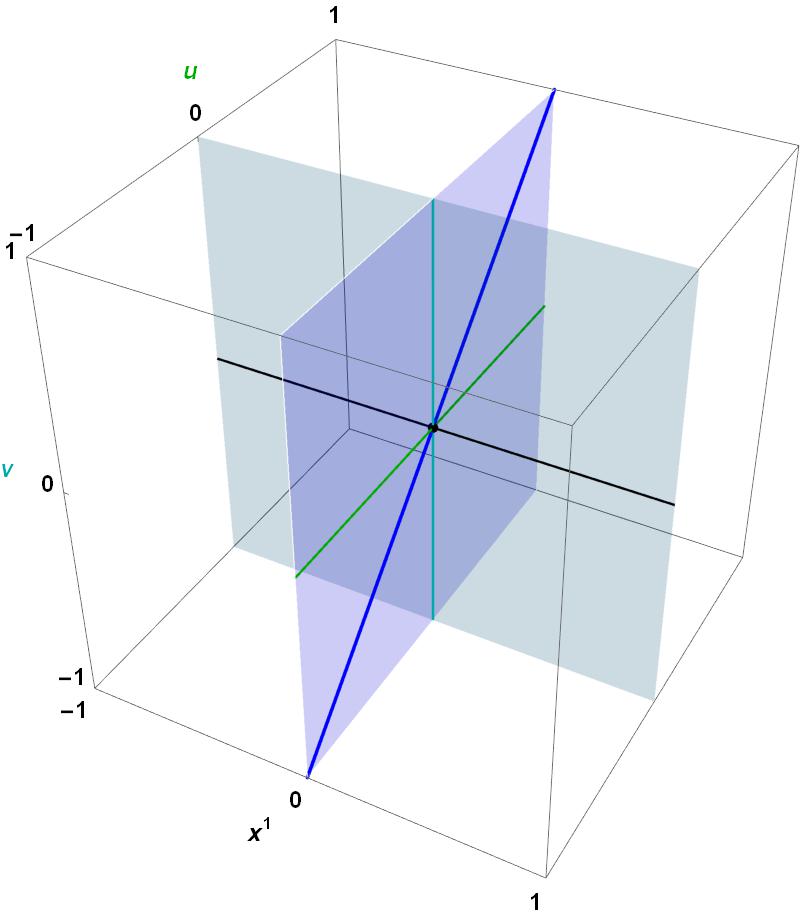}
\\
\vskip-12mm
\null\hskip-8mm
(a)\hskip77mm(b)\vskip-4mm
\caption{{\it The geodesic motion is determined by the conserved quantities. For $\bp=(1,0)$ and $\bk=0$, for example, the  trajectories remain in the hyperplane $x^2=0$ which can therefore be ignored, yielding $3D$ pictures for $\big(x^1(u),u,v(u)\big)$. All  trajectories can be ``straightened out'' by a suitable action of the Carroll group. We took $\alpha(u)=u$.} }
\end{center}%
\label{trajectory}
\end{figure}

Let us mention that choosing instead of (\ref{Pter}):
\begin{equation}
P(u)=
\chi(u)\left(
\begin{array}{cc}
\cosh\beta(u)&\sinh\beta(u)\\
\sinh\beta(u)&\cosh\beta(u)
\end{array}
\right)
\label{Pquater}
\end{equation}
with $\beta(u)$ an otherwise arbitrary function, we find
that the associated metric (\ref{nopotform}) is Ricci-flat if
$\ddot{\chi}+\dot{\beta}^2\,\chi=0$. The pp-wave profile becomes
\begin{equation}
K(U)_{ij}X^iX^j=\cB(U)\,X^1 X^2,
\qquad
\cB=\frac{2}{\chi^2}\,\frac{\ d}{du}\left(\chi^2\,\frac{d{\beta}}{du}\right).
\label{HXXbis}
\end{equation}
The examples (\ref{HXX}) and (\ref{HXXbis}) can be combined,  keeping $\alpha$ and $\beta$ independent, to end up with the most general profile.

\item
So far, the functions $\cA$ and $\cB$ encoding the polarization states of the gravitational wave have been traded as arbitrary. Now in the \textit{periodic} case the isometry group is actually $6$-dimensional \cite{exactsol}. In fact, if 
\begin{equation}
K_{ij}(U)X^iX^j
=
\cos(\omega{}U)\,X^1X^2+\half\sin(\omega{}U)\,\left[(X^1)^2-(X^2)^2\right]
\label{Kperiodic}
\end{equation}
with $\omega=\const$, then, putting $Z=X^1+iX^2$, one finds that the advanced time translations 
\begin{equation}
Z\to{}e^{-\half i\omega e}Z,
\qquad
U\to{}U+e,
\qquad
V\to{}V,
\label{TimeTranslations}
\end{equation}
act isometrically for all $e\in\bbR$. This subgroup of the Bargmann group (\ref{Barggroup}) is however clearly not a subgroup of the Carroll group $C(2+1)$ which leaves $U$ fixed, see (\ref{CarinB}). 

\end{enumerate}

\section{Carroll symmetry of an isotropic oscillator}\label{Section:OH}

When $K=-\omega^2\,\Id$ with $\omega=\const$, the metric (\ref{planewave}) does not solve the vacuum Einstein equations and is therefore 
 \emph{not} that of a gravitational wave; it  describes instead an \emph{isotropic harmonic oscillator} \cite{DBKP,DGH91}. However the procedure described in Sec. \ref{isos} can be carried out at once~: $P=\cos(\omega U)\,\Id$ is a solution of (\ref{SL+cond}) 
 ; then (\ref{BJRcoord}) yields  (\ref{nopotform}) with $\rg_{ij}(u)=\cos^2(\omega u)\delta_{ij}$. 
Integrating (\ref{Hmatrix}) we get,
\beq
H(u)=\left(\frac{\tan(\omega u)}{\omega}-\frac{\tan(\omega u_0)}{\omega}\right)\Id.
\label{osciH}
\eeq
Then (\ref{genCarr}) yields the Carroll action on  oscillator-Bargmann space.
Redefining the time and renaming $\bx$ and $v$ allows us to recover, moreover, Niederer's transformation \cite{Niederer,BDP},
\beq
t=\frac{\tan(\omega U)}{\omega},
\qquad
\br\equiv \bx=\frac{\bX}{\cos(\omega U)},
\qquad
s\equiv v=V-\frac{\omega\, \bX^2}2\tan(\omega U),
\label{Ntrafo}
\eeq
which maps every (half-oscillator-period)$\times \bbR^{d+1} $ conformally onto the Bargmann space, $ \bbR^{d+1,1}, $ of a free particle, 
$
d{\br}^2+2dt\,ds =
{\cos^{-2}(\omega U)}
\big(d\bX^2+2dU\,dV-\omega^2{\bX}^2\,dU^2\big).
$
It is worth noting that, in terms of the redefined time, (\ref{osciH}) is in fact 
$
H(t)=(t-t_0)\,\Id,
$ 
consistently with what we had found in the flat case.

We mention that the Niederer trick (\ref{Ntrafo}) could be used for an alternative derivation of the Carroll symmetry. Recall first that the free (Minkowski) metric  has the (extended) Schr\"odinger group as group of $\xi$-preserving conformal symmetries \cite{DGH91}. The latter is in turn exported to the oscillator by (\ref{Ntrafo}).   
 Now, as said before, the $t=0$ slice of flat Bargmann space is a Carroll manifold $\cC^{d+1}$, with the Carroll symmetry group embedded into the Bargmann group by eliminating the time translations (\ref{Barggroup}).
 Moreover, for $t=0$\footnote{For $t=t_0=\const$ the conformal factor is constant and can be absorbed by a redefinition of the metric.} 
the conformal factor is equal to one. The image is therefore an isometry.  
However, the Niederer trick fails to work in the anisotropic case (and thus for a gravitational wave), whose Bargmann space is not conformally flat \cite{ZHAGK}.

\section{Scattering of light by a gravitational wave}\label{Section:AdditionalComments}

We conclude by some additional comments on the scattering of light by a plane gravitational wave.
In \cite{Westerberg:2014lra} the analogy between
photon  production in a medium with a time-dependent refractive index and
particle production (and its absence)
\cite{G_Rosen,Gibb75,Deser:1875zz,Liu:2007yk,Jones:2016zqw}
by gravitational waves 
has been stressed and the prospects for
laboratory experiments explored. For this purpose
the Baldwin-Jeffery-Rosen form of the metric would seem to be the most useful.
In fact one may use the general formulae
developed by Tamm, Skrotskii and Plebanski
\cite{Tamm,Skrotskii,Plebanski,McCall} to extract the relevant
permittivities $\epsilon^{ab}= \epsilon^{ba}$  and
permeabilities $\mu^{ab}=\mu^{ba} $.
We mainly follow the notation 
of \cite{Plebanski} but use the opposite signature convention
and a different notation for permittivity and permeability.

\goodbreak

We define $t=x^0$, $z=x^3$,
$ 
u=\frac{1}{\sqrt2}(z-t),\,  v=\frac{1}{\sqrt2}(z+t)\,
$  
and write the metric~(\ref{nopotform}) as
\begin{equation}
\rg
=-dt^2 + g_{ab}(u)dx^adx^b,
\qquad
g_{ab}\,dx^adx^b = a_{ij}(u) dx^i dx^j+dz^2,
\end{equation}
where $a,b=1,2,3$ and  $i=1,2$.
Now in the coordinates $(t,\bx)$, where $\bx=(x^a)$, Maxwell's equations take the usual flat-space form in a medium
\begin{subequations}
\begin{align}
{\rm curl}\, {\bf E} &= -\frac{\partial{\bf B}}{\partial t}
     \,,\qquad {\rm div}\, {\bf B}=0\,,
\\[4pt] 
{\rm curl}\, {\bf H} &= +\frac{\partial {\bf D}} {\partial  t} \,,\qquad {\rm div}\, {\bf D}=0\,,
\end{align}
\label{vacMax}
\end{subequations}
where 
$ 
D^a= \epsilon^{ab}E_b, 
\,
B^a= \mu^{ab}H_b.
$ 
Since $ g_{00}=-1$ and  $g_{0a}=0$ in these coordinates,
there are no magneto-electric effects, which in turn  implies
that the medium is ``impedance matched'', that is,  
$ 
\epsilon^{ab}= \mu^{ab}  = \sqrt{-g}\,g^{ab}.
$ 
In detail, one has
\begin{equation}
  \epsilon^{ij} = \sqrt{\det a(u)}\,\left(a(u)^{-1}\right)^{ij},
  \qquad \epsilon^{33}= \sqrt{\det a(u)}, \qquad \epsilon^{3i}=0.
\end{equation}

If  instead  of the Maxwell equations one were to  look at the Dirac equation, 
one might be able to treat the analogue for gravitational waves of the 
Kapitza-Dirac effect \cite{Kapitza} for electromagnetic waves.

\goodbreak 

\section{Conclusion}

Plane gravitational waves have long been known to have a $5$-dimensional isometry group. The first three parameters are readily identified as translations in transverse space, resp. along one of the light-cone generators. The other two have remained somewhat mysterious, though; here we identify the latter as \emph{Carroll boosts}. In fact, we show that the isometry group of  (\ref{planewave}) is  the  {Carroll group without rotations}.
If the matrix $a=(a_{ij}(u))$ in (\ref{nopotform}) happens  to depend on $u$ periodically, then $u$-translational symmetry is restored and the isometry group is enhanced to a 6-parameter group; see (\ref{TimeTranslations}). 
In the isotropic case, as in section \ref{Section:OH}, rotations are also restored and we end  up with the 7-parameter centrally extended Newton-Hooke group. 
In the flat case, $K\equiv0$, the latter becomes the Bargmann group, which is a subgroup of the $10$-parameter Poincar\'e group.


The appearance of these typically non-relativistic structures is  surprising in that the theory is fully general relativistic. It is also unexpected, since the Carroll group has originally been defined as an ultra-relativistic contraction of the Poincar\'e group. Moreover, bearing in mind that (conformal) Carroll structures have been shown to dwell in the edge of space-time, e.g., conformal infinity of certain solutions of Einstein's equation \cite{DGH-BMS}, it is remarkable to witness the Carroll group appearance in the \emph{bulk} of some instances of the latter, namely gravitational plane waves.
From our point of view, the key formulae are
 
\benu
\item
The metric
in Baldwin-Jeffery-Rosen coordinates, (\ref{nopotform});

\vskip-3mm
\item The exact equation for
the matrix $a$ in (\ref{Ricci=0}) related to various profiles,
cf. Fig.\ref{saddles};

\item The explicit form of the action of the Carroll group, eqn. (\ref{genCarr}), viewed as a subgroup of the Bargmann group, (\ref{CarinB});

\item In BJR coordinates the geodesics are expressed in terms of the conserved quantities associated with the Carroll symmetry through Noether's theorem, (\ref{CarGeo}), and can be determined explicitly when  the matrix
$H(u)$ in (\ref{Hmatrix}) is calculated, yielding Fig.\ref{trajectory}. This fact plays a key r\^ole in our subsequent applications to the memory effect \cite{Memory} and thus underlines the importance of Carroll symmetry for the study of gravitational waves.

\eenu

Our clue is the double role played by  ``Bargmann space'' --- which is both a relativistic space-time and a convenient tool to study non-relativistic physics in one lower dimension. From the group theory point of view, our finding corresponds to the fact that the Bargmann (centrally extended Galilei) and Carroll groups are both subgroups of the Poincar\'e group in one higher dimension -- a way of seeing we find more convenient than the original derivation by group contraction \cite{Leblond,SenGupta,Carrollvs}.
 This is the point of view espoused systematically in Sections \ref{planewaves}, \ref{isos}, and \ref{Section:Examples} to unveil the Carroll structure of the group of isometries of the gravitational waves under study.
  A similar argument, developed in Section \ref{Section:OH}, 
  sheds some light  also on the Niederer trick \cite{Niederer}.  
Section \ref{Section:AdditionalComments} comments about using BJR coordinates in studying the scattering of light by gravitational waves.

In this paper we identified the isometries of plane gravitational waves with the Carroll group with no rotations.  
Conformal extensions can also be studied, through, with the remarkable outcome that the celebrated BMS group is, in fact, a conformal Carroll group \cite{DGH-BMS}. It is puzzling to ask what role (if any) the latter could play for gravitational waves.

%
\begin{acknowledgments} 
Enlightening discussions with T. Sch\"ucker are warmly acknowledged. GWG would like to thank the 
{\it Laboratoire de Math\'ematiques et de Physique Th\'eorique de l'Universit\'e de Tours}  for hospitality and the  {\it R\'egion Centre} for a \emph{``Le Studium''} research professor\-ship. PH is grateful  
 for hospitality at the IMP of the Chinese Academy of Sciences in Lanzhou. Support by the National Natural Science Foundation of China (Grant No. 11575254) is acknowledged.
\end{acknowledgments}
\goodbreak



\end{document}